\documentclass[aps,prd,amsmath,amssymb,preprintnumbers]{revtex4}
\usepackage{graphicx}
\usepackage{color}

\newcommand{\bmp}{\noindent\begin{minipage}{16cm}}
\newcommand{\emp}{\end{minipage}\vskip 7mm} 

\usepackage{bm}
\usepackage{bbm}

\newcommand{\beq}{\begin{equation}}
\newcommand{\eeq}{\end{equation}}
\newcommand{\bea}{\begin{eqnarray}}
\newcommand{\eea}{\end{eqnarray}}
\newcommand{\ba}{\begin{array}}
\newcommand{\ea}{\end{array}}
\newcommand{\bi}{\begin{itemize}}
\newcommand{\ei}{\end{itemize}}
\newcommand{\bn}{\begin{enumerate}}
\newcommand{\en}{\end{enumerate}}
\newcommand{\bc}{\begin{center}}
\newcommand{\ec}{\end{center}}

\newcommand{\gsim}{\lower.7ex\hbox{$\;\stackrel{\textstyle>}{\sim}\;$}}
\newcommand{\lsim}{\lower.7ex\hbox{$\;\stackrel{\textstyle<}{\sim}\;$}}

\definecolor{rossoCP3}{cmyk}{0,.88,.77,.40}

\begin{document}
\title{\Large  \color{rossoCP3}  
Charge Asymmetric Cosmic Rays \\ as a probe of \\ Flavor Violating Asymmetric Dark Matter} 
\author{Isabella {\sc  Masina}$^{\color{rossoCP3}{\clubsuit},{\heartsuit}}$}
\email{masina@fe.infn.it} 
\author{Francesco {\sc Sannino}$^{\color{rossoCP3}{\heartsuit}}$}
\email{sannino@cp3-origins.net} 
\affiliation{
$^{\color{rossoCP3}{\clubsuit}}${\mbox{Dip.~di Fisica dell'Universit\`a di Ferrara and INFN Sez.~di Ferrara, 
Via Saragat 1, I-44100 Ferrara, Italy}}}
\affiliation{
$^{\color{rossoCP3}{\heartsuit}}${\mbox{CP$^{ \bf 3}$-Origins and DIAS, 
University of Southern Denmark,  Campusvej 55, DK-5230 Odense M, Denmark}}}
\begin{abstract}
The recently introduced cosmic sum rules combine the data from PAMELA and Fermi-LAT cosmic ray 
experiments in a way that permits to neatly investigate whether the experimentally observed lepton 
excesses violate charge symmetry.  One can in a simple way determine universal properties of the unknown 
component of the cosmic rays. Here we attribute a potential charge asymmetry to the dark sector. 
In particular we provide models of asymmetric dark matter able to produce charge asymmetric cosmic rays. 
We consider spin zero, spin one and spin one-half decaying dark matter candidates. 
We show that lepton flavor violation and asymmetric dark matter are both required 
to have a charge asymmetry in the cosmic ray lepton excesses. 
Therefore, an experimental evidence of charge asymmetry in the cosmic ray lepton excesses 
implies that dark matter is asymmetric. 
 \\[.1cm]
{\footnotesize  \it Preprint: CP$^3$-Origins-2011-21 \& DIAS-2011-07}
\end{abstract}

\maketitle

\vskip 1cm

\section{Charge asymmetry in cosmic rays}

Shedding light on astrophysical and particle physics origins of cosmic rays can lead to breakthroughs in 
our understanding of the fundamental laws ruling the universe. 
In \cite{Frandsen:2010mr} we introduced a new model independent approach for efficiently combine observations 
from different cosmic rays experiments. These {\it cosmic sum rules} showed how to investigate the
possible charge asymmetries in the {\it unknown} components of high energy cosmic rays. We found that 
at present even large deviations from charge symmetry are experimentally viable. 
Interestingly, future experimental observations could better constrain the tolerated amount of asymmetry.  
We now review the sum rule method proposed in \cite{Frandsen:2010mr}.

The data recently collected by PAMELA \cite{Adriani:2008zr} indicate that there is a positron  
excess in the cosmic ray energy spectrum above $10$ GeV.
{The} rising behavior {observed by PAMELA} does not fit previous estimates of the cosmic ray formation 
and propagation implying the possible existence of a direct excess of cosmic ray {positrons} of unknown origins. 
Interestingly PAMELA's data show a clear feature of {such a positron excess but no excess in the anti-protons}
\cite{Adriani:2010rc}. 
{While ATIC \cite{:2008zzr} and PPB-BETS \cite{Yoshida:2008zzc} reported unexpected
structure in the all-electron spectrum in the range $100$~GeV- $1$~TeV, the
picture has changed with the higher-statistics measurements by Fermi-LAT \cite{Abdo:2009zk} and 
HESS \cite{Aharonian:2008aa}, leading to a possible slight additional
unknown component in the CR $e^{\pm}$ flux over and above the standard astrophysical model predictions, 
like for instance the specific Moskalenko and Strong \cite{Strong:1998pw, Baltz:1998xv} one. 
These interesting features have drawn much attention, and many explanations have been proposed: {} For example,
these excesses could be due to an inadequate account of the cosmic ray astrophysical background 
in previous modeling; They could be due to the presence of new astrophysical sources;
They could also originate from annihilations and/or decays of dark matter. 
We refer to \cite{Fan:2010yq} for a recent review.  

Whatever the origin of these excesses might be, the observed flux of electrons and positrons can be written as 
the sum of two contributions:
A background component, $\phi_{\pm}^B$, describing all known astrophysical sources (which, at least for the 
electrons, are considered to be quite well known), and an unknown component, $\phi_{\pm}^U$. 
Explicitly:
\beq
\phi_\pm = \phi_\pm^U + \phi_\pm^B  \ .
\eeq
The component $\phi_{\pm}^U$ is the one needed to explain the features in the spectra 
observed by PAMELA and Fermi-LAT. These experiments measure respectively the positron fraction and 
the total electron and positron fluxes as a function of the energy $E$ of the detected $e^\pm$, i.e.: 
\beq
P(E) = \frac{\phi_+(E)}{\phi_+(E) + \phi_-(E)}\ , \qquad  F(E) = \phi_+(E) + \phi_-(E) \ .
\eeq
The left-hand side of the equations above refer to the experimental measures. 

Our aim is to investigate the unknown contribution leading to the lepton excesses in cosmic rays. 
Clearly, this can be done only assuming a definite astrophysical background model. 
In particular we want to investigate the fundamental properties of the unknown contribution, 
like its charge asymmetry. This step is necessary to better understand its origin. 
The contribution from the unknown source can be written as: 
\begin{eqnarray}
\phi_+^U(E) & = &  P(E)~F(E) -\phi_+^B(E) \ ,  \\ 
\phi_-^U(E) & = & F(E)~ \left(1-P(E)\right) -\phi_-^B(E) \ .
\end{eqnarray}

We model the background spectrum using 
\beq
\phi_\pm^B(E)=N_B B^\pm(E)~~, 
\eeq
where $N_{B}$ is a normalization coefficient and $B^\pm(E)$ are provided using specific astrophysical models. 
In this paper we adopt the popular Moskalenko and Strong model \cite{Strong:1998pw, Baltz:1998xv},
for which 
$B^\pm(E)$ are given, for example, in \cite{Frandsen:2010mr}.
We checked that our results remain unchanged when using instead the astrophysical background model 
adopted by the Fermi-LAT Collaboration (model zero) \cite{Grasso:2009ma, Ibarra:2009dr}.

In terms of their sum:
\begin{equation}
\phi_+^U(E) +\phi_-^U(E) =  F(E)- (\phi_-^B(E)+\phi_+^B(E))   \ . 
\label{sumrules}
\end{equation}
The latter equation implies $F(E)/(B^-(E)+B^+(E))\geq N_B$.
 
The ratio of the unknown fluxes is thus a direct measure of the charge asymmetry of the source 
of the high energy cosmic rays \cite{Frandsen:2010mr}: 
\beq
r_U(E) \equiv \frac{\phi_-^U(E)}{\phi_+^U(E)}= \frac{F(E) ~(1-P(E))-\phi_-^B(E)}{P(E)~F(E)-\phi_+^B(E)}~~.
\label{ruu}
\eeq

This equation can be rewritten as
\beq
R(E) \equiv \frac{F(E)}{B^-(E)} ~\frac{ 1-(1+r_U(E)) P(E)}{1-r_U(E) \frac{\phi_+^B(E)}{\phi_-^B(E)}} = N_B \ .
\label{sumrulefinal}
\eeq
Although the sum rule $R(E)$ seems to depend on the energy it should, in fact, be a constant as is clear 
from the right hand side of the previous equation.  
{This leads to a nontrivial constraint linking together in an explicit form the experimental results, 
the model of the backgrounds and the dependence on the energy of the unknown component charge asymmetry.}  
Since we use simultaneously the results of Fermi-LAT and PAMELA we consider the common 
energy range, {\i.e.} from about $20$ to $100$ GeV. 

In order to test whether current data could support charge asymmetric cosmic rays in \cite{Frandsen:2010mr} 
we considered the oversimplifying assumption that $r_U$ is nearly constant as function of the energy, 
in the energy region common to PAMELA and Fermi-LAT.  
We then extrapolated the prediction for the positron fraction and compared it with the 
PAMELA results \cite{Frandsen:2010mr}. We summarize our results in 
Fig.~\ref{predictionP}.
\begin{figure}[h!]
\begin{center} 
\includegraphics[width=8cm]{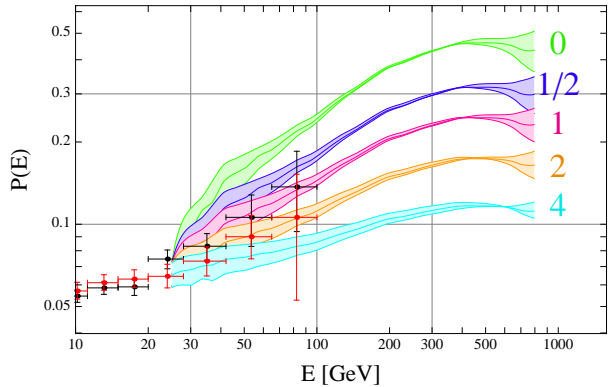} 
\end{center}
\vspace*{-0.5cm} 
\caption{Positron fraction $P(E)$ as a function of the energy $E$ 
of electrons and positrons for different values of $r_U=0,1/2,1,2,4$, from top to bottom.
Secondaries are estimated according to the expressions in \cite{Strong:1998pw, Baltz:1998xv}. The shaded regions reflect the one sigma errors of Fermi-LAT and PAMELA when determining the allowed values of $N_B$ deduced from the sum rule \eqref{sumrulefinal}. We display the 2010 (lower red) and 2008 (upper black) PAMELA data, with 1$\sigma$ error bars.}
\label{predictionP}
\end{figure}
It is clear that the resulting picture shows that the current data allow for a substantial 
charge asymmetry corresponding to $r_U \neq 1$.  

The special case $r_U=1$ is an automatic prediction of a great deal of models for dark matter which 
assume that charge symmetry holds both in the production and propagation of the unknown component of the 
cosmic rays. Symmetric dark matter always implies $r_U=1$, as we will prove in the following.

\section{Flavor Violating Asymmetric Dark Matter Models}

In this work we attribute the experimental cosmic rays excesses to the existence of a dark matter component 
of asymmetric type. This type of candidates appeared first in \cite{Nussinov:1985xr} in the form of technibaryons, and in \cite{Gudnason:2006ug} as Goldstone bosons. Since then, asymmetric dark matter candidates of all types have appeared in the literature ~\cite{Gudnason:2006ug, Foadi:2008qv,Khlopov:2008ty,Dietrich:2006cm,Sannino:2009za,Ryttov:2008xe,Kaplan:2009ag,Frandsen:2009mi,Belyaev:2010kp}.
We should note that the possibility of mixed dark matter with a thermally produced symmetric component and an asymmetric component \cite{Belyaev:2010kp}. Having reviewed the constrains coming from the cosmic sum rules on the cosmic rays 
charge asymmetry \cite{Frandsen:2010mr} we now construct models able to naturally account for such an asymmetry. 
This will also allow us to determine the explicit energy dependence of $r_U(E)$. 
 
Furthermore we will  show that violation of the flavor symmetry in the dark sector is essential to have 
phenomenologically viable charge asymmetries. Flavor violating operators from the dark sector were 
discovered in \cite{DelNobile:2011uf}. 

We will consider here three distinct types of dark matter. 
A complex scalar $T$, a spin-one state $T_{\mu}$ and a fermionic one $N$ similar to the 
fourth generation neutrinos. 
We assume that dark matter is of asymmetric type and therefore 
 only half of the decays allowed at the Lagrangian level are dominant, 
i.e. the ones deriving from the surviving component. In this investigation we assume dark matter 
to couple only to leptons. 

\subsubsection{Scalar Asymmetric Dark Matter}

The scalar interactions we consider are therefore: 
\begin{equation}
c_{\ell \ell'}\,T \overline{\ell} \ell' + {\rm h.c.} \ .
\label{SDM}
\end{equation}
Here  $\ell= e, \mu$ or $\tau$ and we assume summation over the standard model leptonic flavor 
indices. A generic $c_{\ell \ell'}$ leads to violations of the lepton numbers. As explained above we assume that, 
during the evolution of the universe, an asymmetry
in the relic densities of $T$ and $T^*$ arises. We further consider the case in which $T^*$ has disappeared 
and that we are left today only with $T$. The latter decays via the first interaction term given in \eqref{SDM}.
Explicitly, this leads to 
\beq
T \rightarrow \ell^-_{L} {\ell'}^+_{L}  + \ell^-_{R} {\ell'}^+_{R} ~~,
\label{TLL}
\eeq
where the notation means that leptons in the pair are produced with the same helicity 
and that there is equal probability for both chiralities. Parity is thus conserved.
If $\ell \neq \ell'$ then asymmetric dark matter implies charge-conjugation violation in the decay.
Clearly, if $\ell$ ($\ell'$) is not directly the flavour $e$, electrons (positrons) are produced
in its decay chain. If we were to have symmetric type dark matter then, as it is clear from \eqref{TLL}, 
we would have an equal energy spectrum of electron and positrons since they would be produced via 
$T$ and $T^{*}$. 

\subsubsection{Vector Asymmetric Dark Matter coupling to a L or R lepton current.}

Similarly we consider the following left and right handed leptonic currents involving the complex $T_{\mu}$ 
spin one massive dark matter field
\begin{equation}
d_{\ell \ell'}\, T_{\mu}  \overline{\ell}  \gamma^{\mu}\frac{1 \pm \gamma_5}{2} \ell' + {\rm h.c.} \ .
\end{equation}

Assuming again asymmetric spin one dark matter component made by the $T_{\mu}$ states we find that the 
decay products are
\begin{eqnarray}
T_{\mu} \rightarrow \ell^-_{L} {\ell'}^+_{R} ~~,~~{\rm{L-current}}\  ,\qquad {\rm and} \qquad 
T_{\mu} \rightarrow \ell^-_{R} {\ell'}^+_{L} ~~,~~{\rm{R-current}} \ .
\end{eqnarray} 
By construction this model violates parity maximally. Furthermore if we have no asymmetry in the dark matter 
relic density there will be no charge asymmetry in the cosmic rays and, last but not the least, flavor violation 
is needed to achieve this charge asymmetry.

\subsubsection{Fermionic Dark Matter with the quantum numbers of a fourth active neutrino.}

We also consider semileptonic decays from a fourth generation like heavy neutrino \cite{Masina:2011ew} 
stemming from interactions of the type:
\begin{equation}
p_{N\ell} \, W_{\mu}^{+} \,\overline{N} \gamma^{\mu} \frac{1-\gamma_5}{2} \ell  + {\rm h.c.} \ ,  
\end{equation}
 with $p_{N\ell} $ the coupling strength. The heavy Dirac neutrino $N$ carries a new lepton number which 
differentiates it from its conjugate $\bar N$. An asymmetry would imply that either $N$ or $\bar N$
are left today to be a fraction of dark matter. Accordingly, only one among these decays take place:
\beq
N \rightarrow \ell^-_{L} W^+_L  ~~,~~~~\bar N \rightarrow \ell^+_{R} W^-_L \ .
\eeq
The labels stress the fact that in both cases the $W$ boson is longitudinally polarized. A symmetric-type 
dark matter would still produce charge symmetric cosmic rays despite the evident flavor violation. 
The Majorana heavy neutrino, as a corollary of the previous statement, would lead to charge symmetric 
cosmic rays \cite{Masina:2011ew}. Recently other models have been explored providing also charge asymmetric cosmic rays \cite{Chang:2011xn}.

\section{Charge Asymmetric Cosmic Rays}

We now discuss the energy spectra of the high energy electrons and positrons coming from the decays of the 
different types of asymmetric dark matter envisioned above. It is possible to investigate the different types 
of dark matter by simply patching together the spectra coming from the dark matter direct products. 
We now consider in turn the different decay products. For definiteness, we adopt here the same propagation model
discussed in detail in our previous work \cite{Masina:2011ew}. This is a propagation model commonly used when 
investigating dark matter as primary source of cosmic ray excesses.

When the flavor of $\ell$ is the electron one the resulting electrons or positrons spectra are monochromatic 
with an energy $E_e=M_{DM}/2$, for the scalar and spin-one dark matter. There is a tiny kinematic correction 
to this value for the semileptonic one \cite{Masina:2011ew}. After propagation they display therefore a hard 
spectrum depicted in fig.~\ref{fig-e3phi} for $M_{DM}=1,3$ and $10$ TeV.

{}For $\ell=\mu$, the initial muon has energy $M_{DM}/2$ and the resulting electrons (positrons) produced 
in the decay of $\mu^-_{L}$ ($\mu^+_{R}$), 
turn out to have a slightly harder spectrum than those produced in the decay of $\mu^-_{R}$ ($\mu^+_{L}$).  
The associated spectral functions after propagation are displayed in fig.~\ref{fig-e3phi}. 
Clearly, for an unpolarized $\mu^\pm$, as it is the case of a scalar dark matter, 
the energy spectrum of $e^\pm$ is given by the mean of the solid and dotted curves. A similar result applies 
to the case of $\ell=\tau$ and the associated spectral curves are  
displayed in fig.~\ref{fig-e3phi}. Finally, in the same figure we show also the spectrum of $e^\pm$ coming 
from the decay of a longitudinally polarized $W^\pm$. {}For the specific details of the propagation model, 
intermediate computations and explicit formulae we refer to \cite{Masina:2011ew}.

\begin{figure}[h!]
\begin{center} 
\includegraphics[width=7cm]{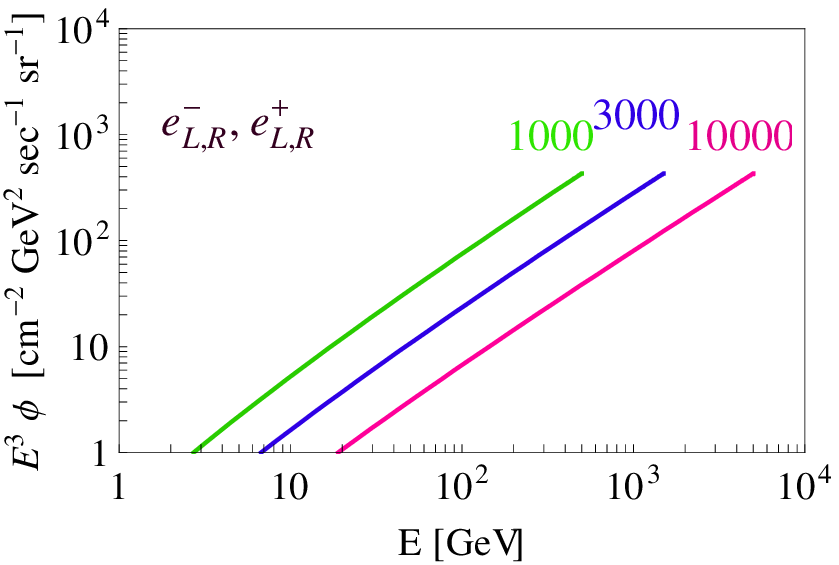} \qquad
\includegraphics[width=7cm]{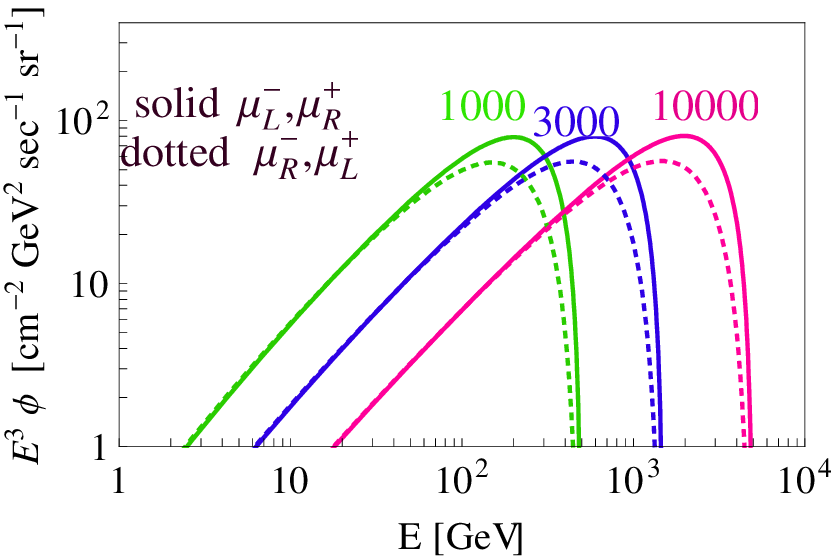} 
\includegraphics[width=7cm]{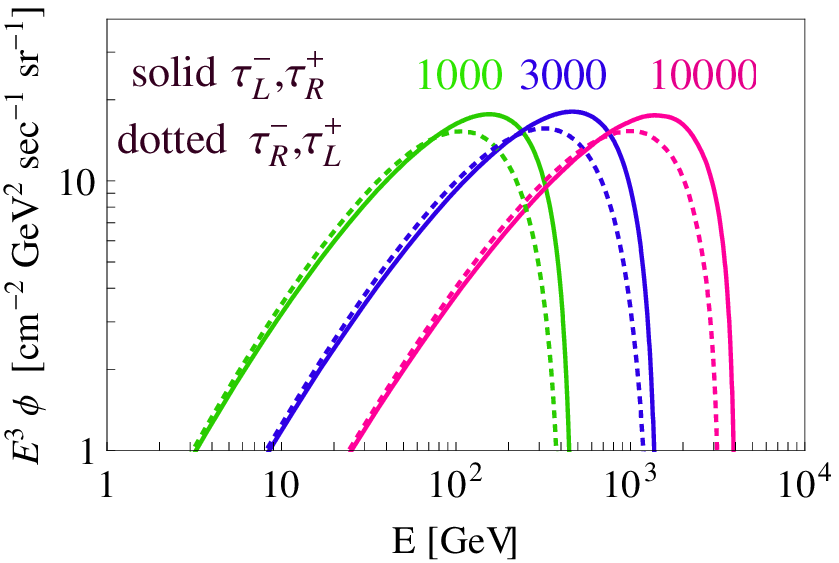} \qquad 
\includegraphics[width=7cm]{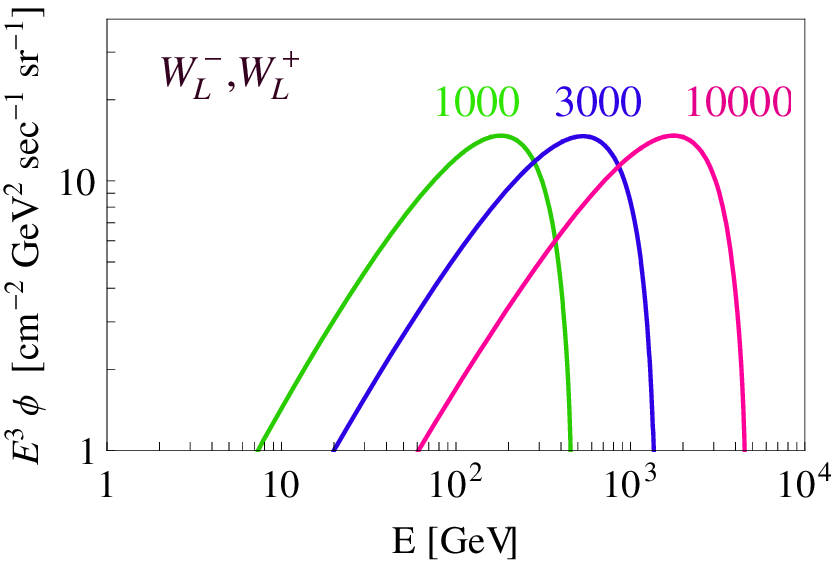} 
\end{center}
\caption{Different spectra for the cosmic rays electrons and positrons coming from the primary particles 
shown in the different panels. We assumed $M_{DM}$ to be respectively $1000$, $3000$ and $10000$ GeV. The lifetimes, fixing the size of the couplings, are chosen to be $10^{22}$~s for the plots.}
\label{fig-e3phi}
\end{figure}

 We are now ready to investigate our predictions for charge asymmetric cosmic rays by computing $r_U(E)$ 
defined in \eqref{ruu} as the ratio of the electron to positron fluxes of the unknown component, here assumed 
to come from asymmetric dark matter, measured at Earth.
 
 \subsubsection{Scalar and Vector Asymmetric Dark Matter prediction for Charge Asymmetric Cosmic Rays}
 
Let us first consider the decay of the scalar and vector dark matter. If the decay is flavour conserving, 
namely $\ell=\ell'$, the electrons and positrons have the same energy spectrum and consequently $r_U=1$ over 
the entire energy range. 

As we explained above a necessary condition for charge asymmetric cosmic rays is the presence of a lepton flavor 
violation. This will lead to a specific energy dependent  $r_U(E)$ different from unity.
For a decaying scalar or vector dark matter there are six possibilities to combine two different 
flavours. Each of these different combinations lead to a specific $r_U(E)$. We report the results for $r_U(E)$ in 
fig.~\ref{fig-ru} for a $3$ TeV decaying dark matter. The decays of a scalar dark matter is represented as the 
inner solid line, the vector
coupling to a L(R)-current is the (dot) dashed curve.  The two panels have been constructed to better elucidate 
the results but carry the same information given that one is the inverse of the other. The (right) left panel 
represents values of $r_U(E)$ mostly (smaller) greater than one.

\begin{figure}[h!] 
\begin{center} 
\includegraphics[width=7cm]{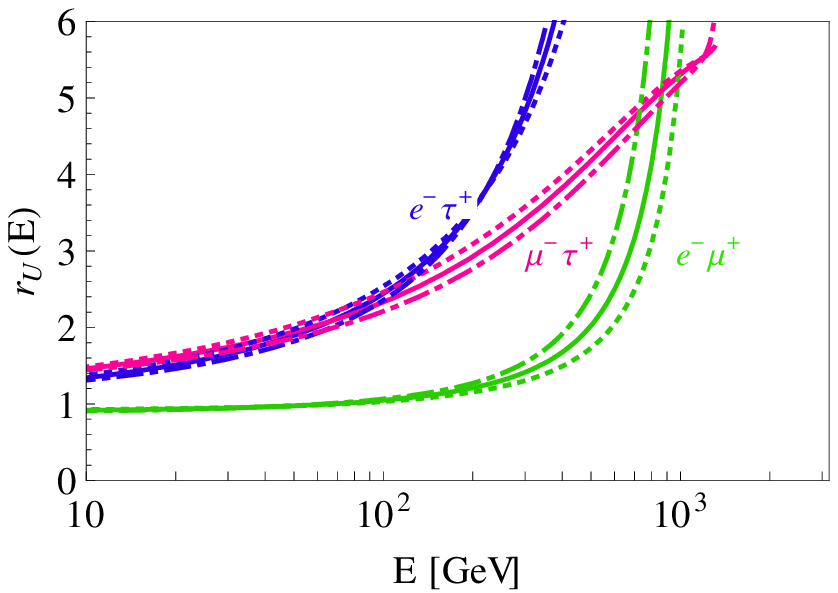} \qquad
\includegraphics[width=7cm]{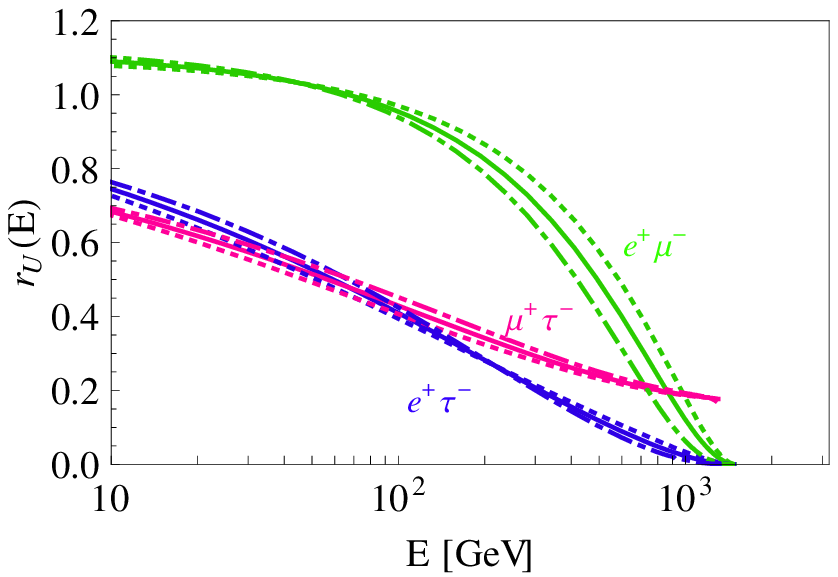} 
\end{center}
\caption{We plot $r_U(E)$ for a $3$ TeV decaying dark matter of either scalar or vector type. The scalar 
dark matter is represented as the inner solid line, the vector dark matter with 
coupling to a L(R)-current is the (dot) dashed curve. }
\label{fig-ru}
\end{figure}

The ratio $r_U(E)$ varies by a factor of two in the energy range for primary $\mu \tau$ leptons. 
The reader will also notice that there is little difference between the scalar and the vector dark matter. 
It is clear that a vector dark matter coupling to the leptons vectorially, i.e. of left plus right type current, 
would be indistinguishable from scalar dark matter.

 \subsubsection{Fermionic Asymmetric Dark Matter prediction for Charge Asymmetric Cosmic Rays}

In the case of a decaying heavy neutrino the two processes are either 
$N\rightarrow W^+ \ell^-$ or $\bar N\rightarrow W^- \ell^+$. If the heavy neutrino is of Majorana type, 
clearly $r_U=1$.
As it is the case for the bosonic asymmetric dark matter we find that lepton flavour must be violated to 
have a $r_U$ which is not unity over the all range of energy. The resulting ratio $r_U(E)$ is displayed in 
fig.~\ref{fig-ruW}.

\begin{figure}[h!]
\begin{center} 
\includegraphics[width=7cm]{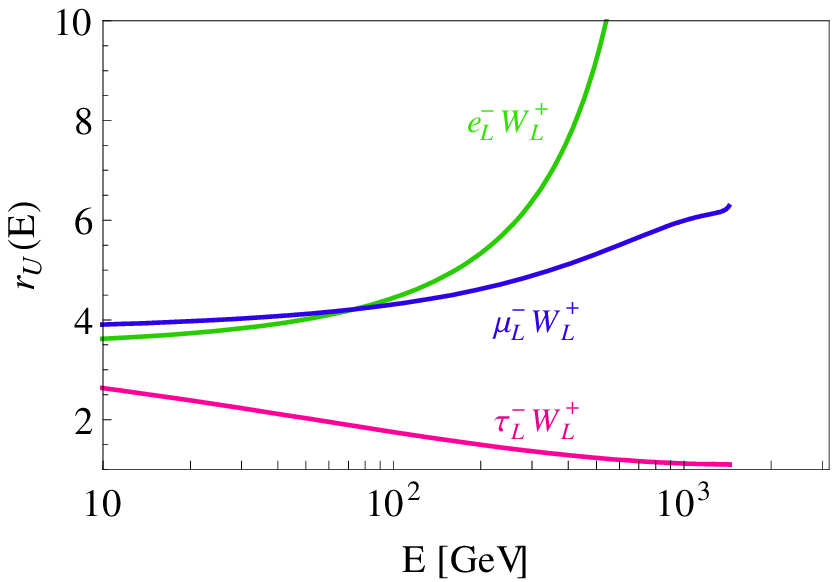} \qquad 
\includegraphics[width=7cm]{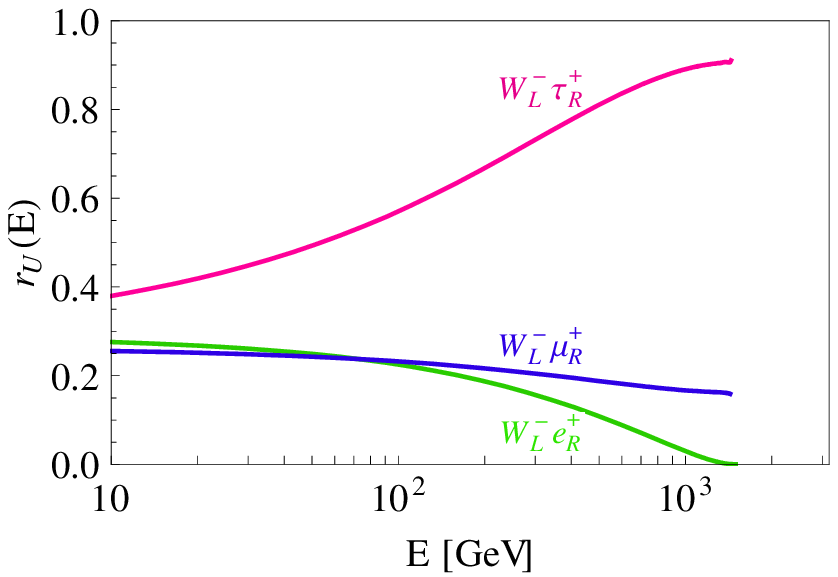} 
\end{center}
\caption{We plot $r_U(E)$ for a $3$ TeV decaying dark matter of Dirac type. The left (right) panel 
corresponds to $N\rightarrow W^+ \ell^-$  ($\bar N\rightarrow W^- \ell^+$).  }
\label{fig-ruW}
\end{figure}

We find that in these decays the ratio $r_U(E)$ is nearly constant or linear in the energy for primary 
$\mu$ and $\tau$ leptons.


\section{Comparison with Fermi-LAT and PAMELA}
 
To connect to previous studies we start by considering the flavour preserving dark matter decays, 
i.e. $\ell=\ell'$ for the scalar and vector dark matter. The results are presented in fig.~\ref{fig-FC}. It is known that $\ell=e$ cannot fit the Fermi-LAT data. 
The case of the muon pair is instead compatible with both Fermi-LAT and PAMELA while the tau pair yields 
an even better fit. The dashed (upper) and dot-dashed (lower) curves, corresponding to a specific primary 
decay mode, represent respectively the vector dark matter coupling with the left and right current. 
Interestingly the difference between the scalar and left or right vector dark matter is within reach 
of the experimental errors. This implies that experiments can determine, in the near future, whether 
the dark side violates parity maximally as it is the case for the bright side. 
\begin{figure}[h!]
\begin{center} 
\includegraphics[width=7cm]{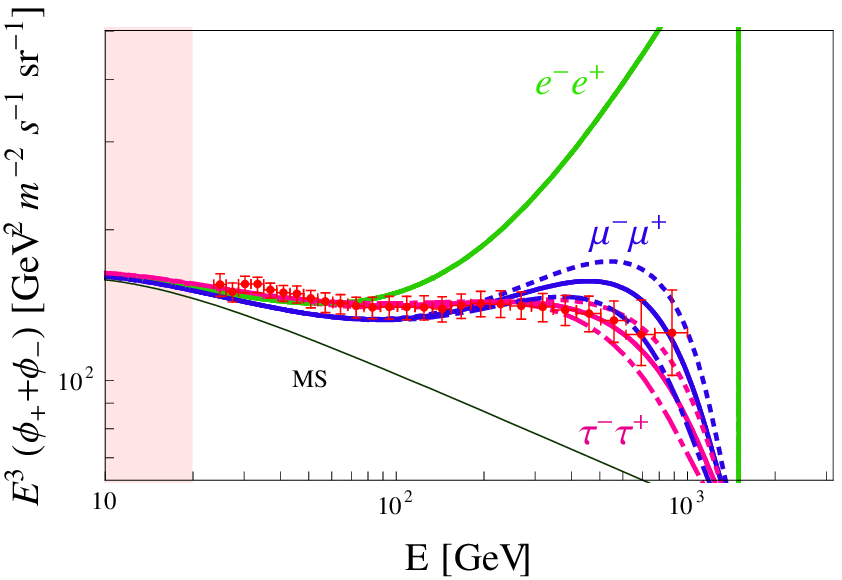}  \qquad 
\includegraphics[width=7cm]{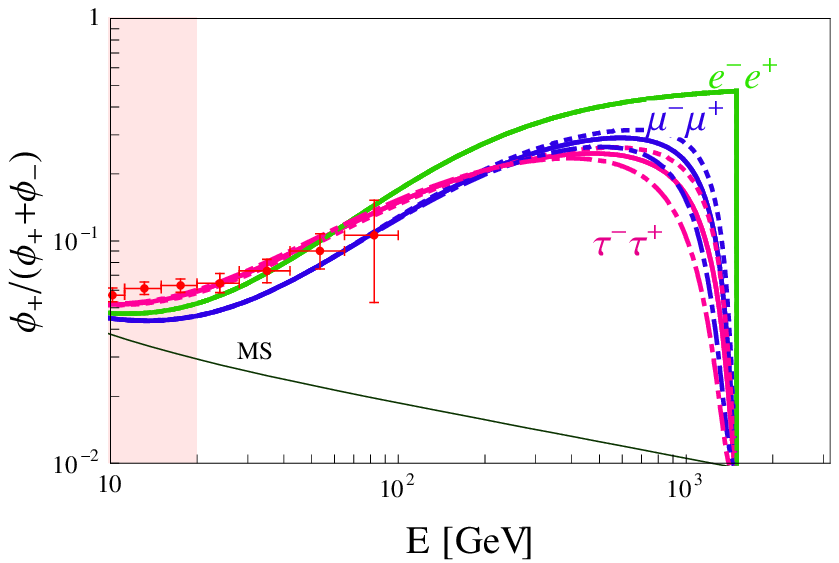} 
\end{center}
\caption{Comparison of flavour preserving dark matter decays for Fermi-LAT (left panel) and PAMELA (right panel).  
The solid lines correspond to scalar decays while the dashed (upper) and dot-dashed (lower) curves, 
corresponding to a specific primary decay mode, represent respectively the vector dark matter coupling with 
the left and right current. The MS stands for the Moskalenko and Strong background. We have chosen the decay 
time for the decaying in $e^+ e^-$ to be $10^{26}$~s, for $\mu^+ \mu^-$ we have taken $1.5 \times 10^{26}$~s, 
and for $\tau^+ \tau^-$ we have taken $0.5 \times 10^{26}$~s. These values correspond to the best fit to the 
Fermi-LAT data. We have also chosen the parameter $N_B = 0.64$. }
\label{fig-FC}
\end{figure}

We now study the flavour violating case in which $\ell \neq \ell'$ for the bosonic and fermionc asymmetric 
dark matter case starting from the bosonic case presented in fig.~\ref{fig-FV}. The comparison with Fermi-LAT 
is shown in the upper panel while the comparison with PAMELA is shown in the two figures of the lower panel.
\begin{figure}[h!]
\begin{center} 
\includegraphics[width=7cm]{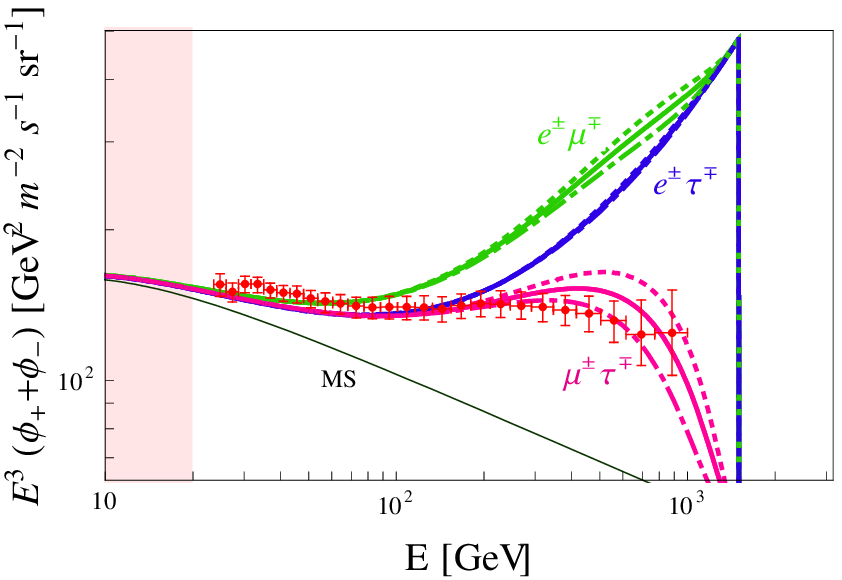} \vskip 1cm
\includegraphics[width=7cm]{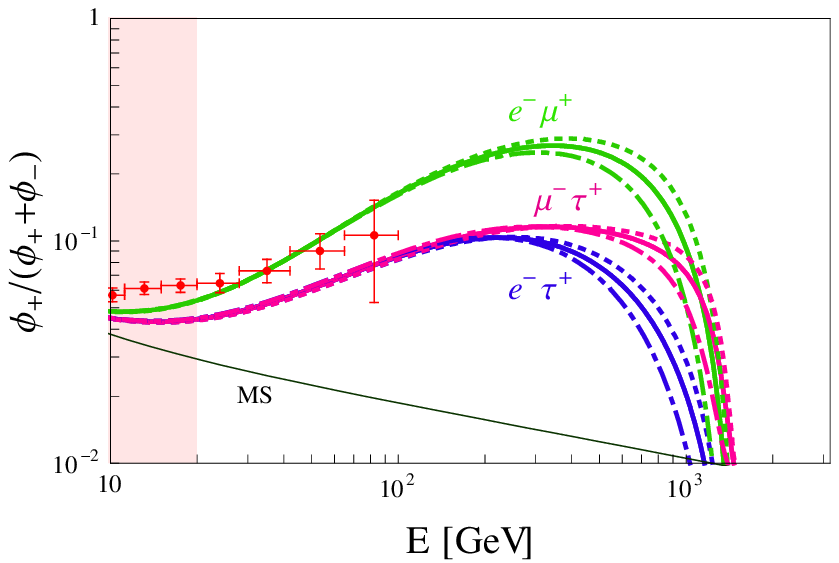}  \qquad 
\includegraphics[width=7cm]{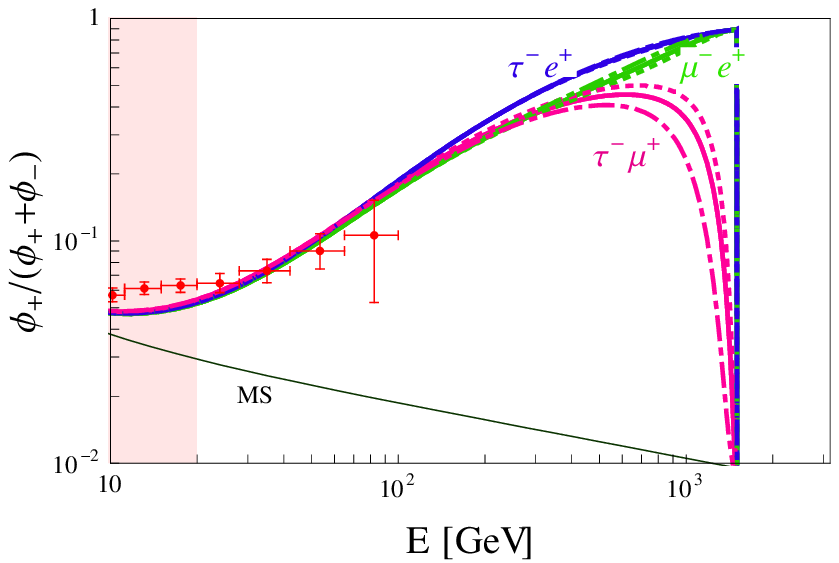} 
\end{center}
\caption{Comparison of flavour violating dark matter decays for Fermi-LAT (upper panel) and PAMELA (lower panels).  The solid lines correspond to scalar decays while the dashed (upper) and dot-dashed (lower) curves, corresponding to a specific primary decay mode, represent respectively the vector dark matter coupling with the left and right current. The MS stands for the Moskalenko and Strong background. We have chosen the lifetime to be $10^{26}$~s for all the primary decays and $N_B = 0.64$.
}
\label{fig-FV}
\end{figure}
  Fermi-LAT results dictates that the most promising processes must not involve electron/positron 
primaries from the decays. When comparing the three pictures there is a preference for the $\mu \tau$ primary 
lepton pair. Furthermore PAMELA type experiment is sensitive to charge asymmetric cosmic rays. 
In particular by reducing the experimental errors and increasing the higher end of the energy range one 
will be able to determine whether the cosmic rays are asymmetric and which kind of asymmetry produces them. 
Interestingly a charge asymmetry stemming solely from  $\mu^- \tau^+$ primaries is currently disfavored 
while the $\mu^+\tau^-$ is favored for both PAMELA and Fermi-LAT.

We have performed a similar comparison for the fermion case. The results are shown in fig.~\ref{fig-FVW}. 
It is evident, by inspection, that the best fit occurs for the decay of $\bar{N}$ in $W$ and a right-handed 
anti-muon or anti-tau. 
\begin{figure}[h!]
\begin{center} 
\includegraphics[width=7cm]{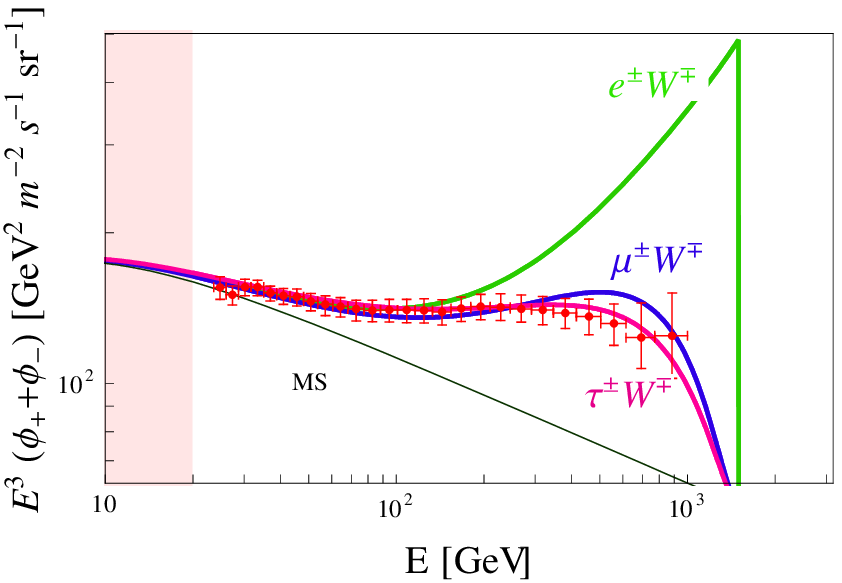} \vskip 1cm 
\includegraphics[width=7cm]{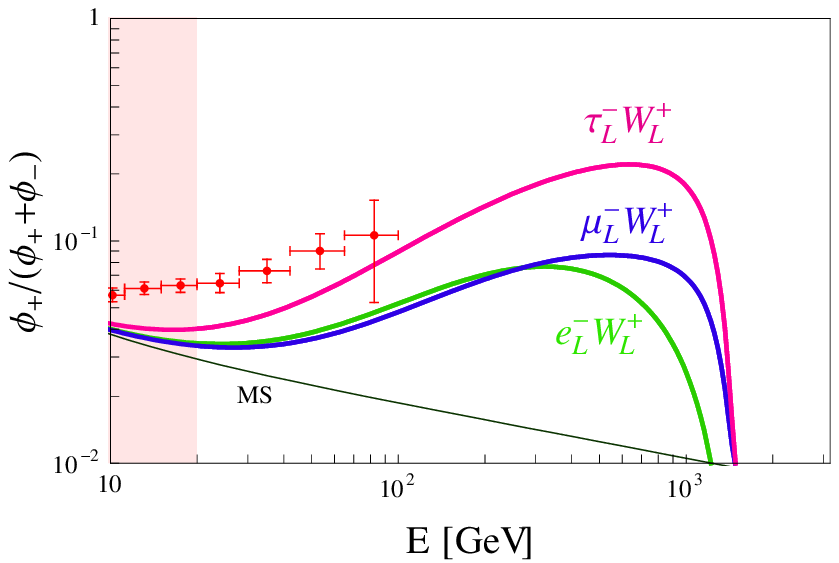}  \qquad 
\includegraphics[width=7cm]{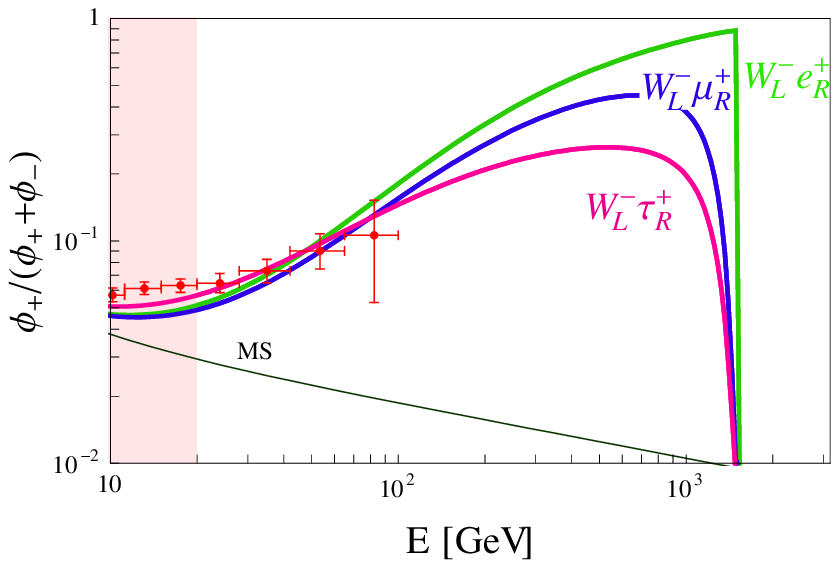} 
\end{center}
\caption{Comparison of flavour violating dark matter decays for Fermi-LAT (upper panel) and PAMELA 
(lower panels) for Dirac type asymmetric dark matter.  The different solid lines correspond to indicated 
primary decay modes. We have chosen the lifetime for decaying in $e W$ to be $10^{26}$~s, for $\mu W$ 
we have taken $1.2 \times 10^{26}$~s, and for $W \tau^-$ we have taken $0.5 \times 10^{26}$~s. We have also 
chosen the parameter $N_B = 0.7$. }
\label{fig-FVW}
\end{figure}
Comparing these processes with the right panel of fig.~\ref{fig-ruW} we find that for the anti-muon 
case there is a stronger, but nearly constant, charge asymmetric component in the cosmic ray while the anti-tau 
process leads to a smaller smaller asymmetry raising towards unity at high energies. The investigation for the 
Majorana case has been performed recently in \cite{Masina:2011ew} and corresponds to $r_U = 1$.  Our results apply also to the case in which the asymmetric dark matter constitutes only a fraction of the dark matter relic density by opportunely modifying the lifetimes. 

Very recent data of the PAMELA collaboration \cite{Adriani:2011xv} on the negative electron flux 
have been released which, however, given the large uncertainties for energies higher than $100$~GeV do not 
affect our results.
 
 For the scalar and vector dark matter there are no constraints coming from the antiproton to 
proton ratio \cite{Adriani:2010rc}. For the heavy neutrino dark matter there is an excess of antiprotons 
which we will compute in the next section. While the interpretation in terms of dark matter annihilations often leads to an unobserved excess of gamma and radio photons, the interpretation in terms of dark matter decays is compatible with photon observations \cite{Nardi:2008ix,Ibarra:2009nw}, even though some channels now start to show some tension \cite{Cirelli:2009dv,Meade:2009iu,Papucci:2009gd,Hutsi:2010ai}.
A discrimination strategy was proposed in \cite{Boehm:2010qt,PalomaresRuiz:2010uu}.  However, the gamma photons analysis, like the one presented in \cite{Chang:2011xn,Cirelli:2010xx} for models similar to ours, show that our results are compatible with the experimental constraints  \cite{Abdo:2010ex,Abdo:2010nz}. The AMS-02 space station experiment will hopefully provide additional relevant informations \cite{ams02}.


\section{Cosmic Ray Antiprotons for the Heavy Neutrino Decay}

In this section we give an estimate for the cosmic ray antiproton flux resulting from the heavy 
neutrino decay and compare it with current data. 
 
Protons and antiprotons are generated via the hadronic decay of the primary $W$ boson, with BR of about $67\%$.
Their energy spectrum is determined by the fragmentation and hadronization processes. In this case an analytic 
approach is not suitable and we rather adopt the numerical recipies provided in ref. \cite{Cirelli:2010xx}.
In particular, the antiproton (proton) flux obtained from a $1.5$ TeV dark matter annihilating 
into $W^+ W^-$ is twice the antiproton (proton) flux in our model.  

The propagation for antiprotons through the galaxy is described by a diffusion equation whose solution 
can be cast in a factorized form as discussed {\it e.g.} in \cite{Cirelli:2010xx}. In this case 
the astrophysical uncertainties
associated to the dark matter profile and to the propagation parameters are large, about one order of magnitude.
We display the antiproton flux in the left panel of fig.~\ref{fig-ap} for the MAX, MED, MIN propagation models
and for $M_{N_1}=3$ TeV and  life time $1.9 \times 10^{26}$~s. The dashed curves show these primary 
antiprotons for MAX, MED, MIN models; the lower shaded region represents the flux of the secondary 
antiprotons according to ref.~\cite{Bringmann:2006im}; the upper three curves correspond to the sum of the 
primary and secondary antiprotons. 
The (lower) PAMELA data \cite{Adriani:2010rc} have smaller error bars with respect to 
the (upper) CAPRICE data \cite{Boezio:2001ac}. 
It turns out that only the MIN propagation parameter set is compatible with the data, the MED one is barely 
compatible, while the MAX seems disfavored.

\begin{figure}[t!]\begin{center}
\includegraphics[width=6.5cm]{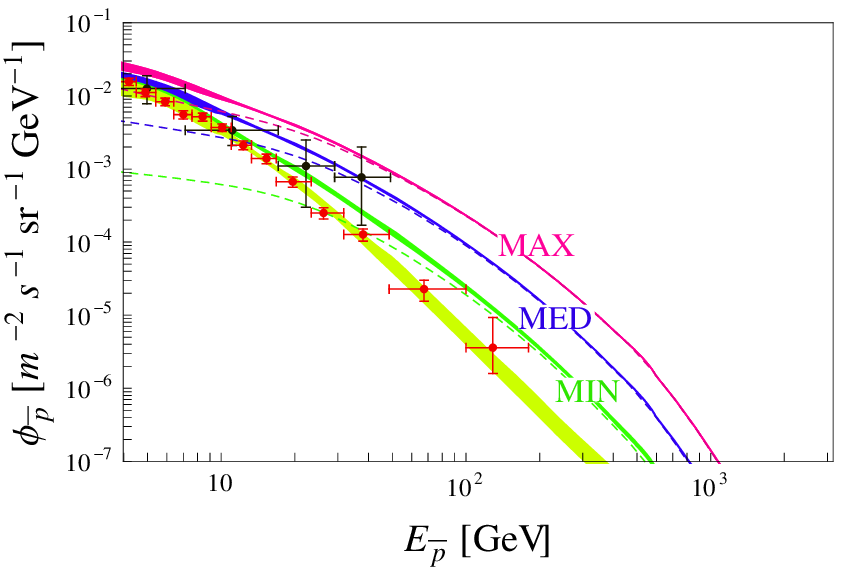}~~~~~~~~\includegraphics[width=6.5cm]{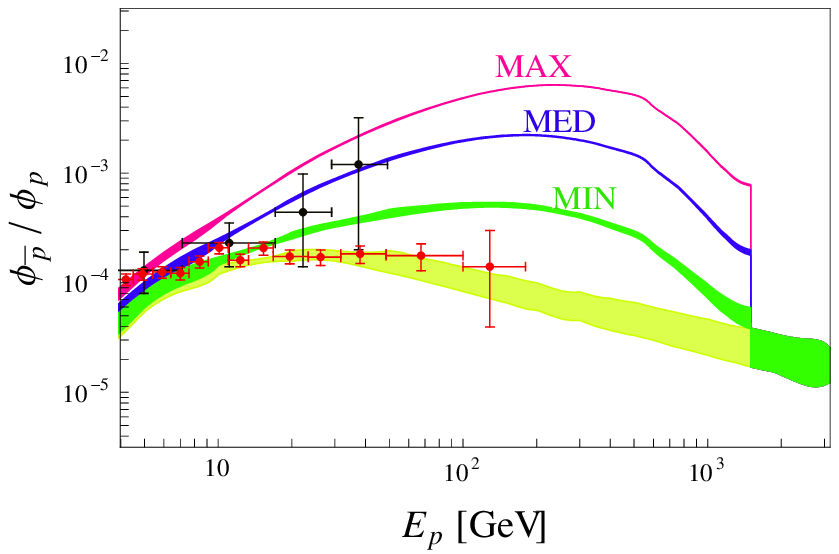}
\end{center}
\vspace*{-0.8cm} 
\caption{Antiproton flux (left) and antiproton/proton ratio (right) for the MAX, MED, MIN propagation models, 
$M_{N_1}=3$ TeV and  life time $1.9 \times 10^{26}$~s. The dashed curves in the left panel show the primary 
antiprotons for MAX, MED, MIN models. The lower shaded region represents the secondary antiprotons only,
according to ref.~\cite{Bringmann:2006im}. The upper three curves correspond to the sum of the primary and
secondary antiprotons. The (lower) PAMELA data \cite{Adriani:2010rc} have smaller error bars with respect to 
the (upper) CAPRICE data \cite{Boezio:2001ac}. }
\label{fig-ap}
\end{figure}

The antiproton/proton ratio is studied in the right panel of fig.~\ref{fig-ap}. For the proton flux, we 
consistently adopt the parameterization of ref.~\cite{Bringmann:2006im} with spectral index equal to $-2.72$.
This parameterization is valid for energies higher than about $10$ GeV. 
Again, the (lower) PAMELA data points \cite{Adriani:2010rc}
are more precise than the (upper) CAPRICE ones \cite{Boezio:2001ac}. 
The plot shows that the heavy neutrino model displays tension
with the PAMELA data and therefore we expect the scalar model to be favored. Future measurements confirming the PAMELA results could be able to rule out the
heavy decaying neutrino model presented here. 
 

\section{Conclusions}

Concluding, in \cite{Frandsen:2010mr} we asked whether cosmic rays could feature a charge asymmetry. We demonstrated, 
by combining the data from PAMELA and Fermi-LAT via sum rules,  that such a a charge asymmetry is experimentally 
viable and can be tested. In   \cite{Frandsen:2010mr}  we did not make any assumption on the specific 
model which could lead to such an asymmetry in the cosmic rays. 
Here we attributed the cosmic rays potential charge asymmetry to the dark sector and provided relevant examples 
of asymmetric, flavor violating, decaying dark matter made of a complex scalar, vector or a Dirac fermion. 
We determined the associated energy dependent charge asymmetry $r_U(E)$ for these models. We then compared our predictions for the charge asymmetric cosmic rays with the data coming from  
PAMELA and Fermi-LAT. 
 
We discovered that dark matter must both be of asymmetric type and violate lepton flavour to generate charge asymmetric cosmic rays excesses.  Therefore a model independent way 
to directly demonstrate that dark matter is of asymmetric type is to observe a charge asymmetry in the 
cosmic rays.



\end{document}